\documentstyle[preprint,aps,epsf]{revtex}

\begin{document}

\title{Ground State Masses and Binding Energies of the Nucleon, Hyperons 
and Heavy Baryons in a 
Light-Front Model 
\footnote{To appear Brazilian Journal of Physics ({\bf BJP}) (2003).}}
\author{E.F. Suisso$^a$, J. P. B. C. de Melo$^b$ and T. Frederico$^a$}
\address{
$^a$Dep. de F\'\i sica, Instituto Tecnol\'ogico da Aeron\'autica, \\
Centro T\'ecnico Aeroespacial, \\
12.228-900, S\~ao Jos\'e dos Campos, SP, Brazil\\
$^b$Instituto de F\'\i sica Te\'orica, Universidade Estadual Paulista \\
01450-900, S\~ao Paulo, SP, Brazil\\
}
\date{\today}

\maketitle

\begin{abstract}
The ground state masses and binding energies of the  nucleon, 
$\Lambda^0$, $\Lambda^+_c$, $\Lambda^0_b$ are studied within a 
constituent quark QCD-inspired light-front model. The light-front Faddeev 
equations for the $Qqq$ composite spin 1/2 baryons, are derived 
and solved numerically. The experimental data for the masses are 
qualitatively described by a flavor independent effective interaction.
\end{abstract}

%\pacs{12.39Ki,11.10.St,11.10.Gh,12.39.Hg}

%% \begin{multicols}{2}

\section{Introduction}

Modelling the light-front hadron wave-function is a challenge 
as long as the exact wave-function from Quantum Cromodynamics is not 
yet avaliable. The wave-function contains physical information 
complementary to the spectrum.  At scales below 1 GeV, the wave-function 
is described by effective degrees of freedom, and the lowest Fock-state 
component of the light-front hadron wave function
is composed by the minimal number of constituent quarks necessary to 
give the quantum numbers. Within this general framework, 
a light-front QCD-inspired model was recently applied to the 
pion and other mesons \cite{pauli,tobpauli}. A reasonably description of 
the pion structure as well as the masses of the vector and 
pseudo-scalar mesons was found. This model, without confinement, 
describes the vector meson as weakly bound system of constituent 
quarks. The spin does not  play a dynamical role besides justifying 
the contact term coming from the hyperfine interaction. 
In this way the one-gluon-exchange interaction
is simplified to two components only: a contact term and a 
Coulomb-type potential. The contact term is essential to collapse the
constituent quark-antiquark system to form the pion, while the vector meson
is dominated by the Coulomb-type potential. The contact interaction
brings to the model the physical scale of the pion mass which
determines the masses of the other pseudo-scalar and vector 
mesons\cite{tobpauli}. 

Here, we follow closely the work of Ref.\cite{suisso02} and 
revise the extension of the concepts coming from the 
effective QCD-model, applied to mesons \cite{pauli,tobpauli}, 
to study the spin 1/2 low-lying states of the nucleon, $\Lambda ^{0}$, $%
\Lambda _{c}^{+}$ and $\Lambda _{b}^{0}$\cite{suisso02}. We use a 
flavor independent effective interaction between the constituent 
quarks, a property necessary to describe the masses of these 
baryons\cite{suisso02}. The
three-quark relativistic dynamics  of the $Qqq$ system
is formulated within the light-front framework in a
truncated Fock-space\cite{tob92}  which is stable under kinematical boost
transformations\cite{perry} and yields a wave-function covariant under
kinematical boosts \cite{brodsky,karmanov}. We use only the contact
interaction, which in this case  provides to the model the physical scale 
of the mass of the nucleon ground state, while the spin is averaged out. The binding 
energy of the baryon is calculated using three-quark Faddeev 
equations\cite{suisso02} as a function of the mass 
of one of the constituent quarks $(Q)$,  while the bare 
strength of the effective contact interaction and the mass of the quark $q$ 
are kept constant.  This light-front model  with a contact 
force \cite{tob92} has been applied to the proton and described 
its mass, charge radius and electric form factor up to 
2(GeV/c)$^{2}$ \cite{Pacheco95}. Recently, it was also  applied to study
the dissolution of the nucleon at finite temperature and baryonic density 
\cite{beyer01}.

This work is organized as follows. In Sec.II, we present
the coupled integral equations for the Faddeev components of 
the vertex of the three-quark light-front
bound-state wave function from a flavor independent
contact interaction derived in Ref.\cite{suisso02}. 
In Sec. III, we present the numerical results for the masses and binding energies
of the nucleon, $\Lambda ^{0}$, $\Lambda _{c}^{+}$ and $\Lambda _{b}^{0}$, 
obtained from the numerical solution of light-front integral equations.
A key point in this section is the assignement of the constituent  quark 
masses which are done using the experimental
values of the vector meson ground states masses. In Sec. IV, we present
our summary.

\section{Light-front model}

The light-front hyperplane is defined by the time $x^{+}=x^{0}+x^{3}=0$ 
and the position coordinates on the hypersurface are defined
by $x^{-}=x^{0}+x^{3}$ and 
$\vec{x}_{\perp }=(x^{1},x^{2})$ \cite{brodsky,karmanov}. 
The coordinate $x^{+}$ is the light-front time and the  momentum,
$k^{-}=k^{0}-k^{3}$, corresponds to the light-front energy. The 
momentum coordinates $k^{+}$ and $\vec{k}_{\perp }$, are the 
kinematical momenta canonically conjugated to $x^{-}$ and 
$\vec{x}_{\perp }$, respectively. The dynamics of the 
 model is defined by a two-body contact interaction between the
constituent quarks at equal light-front 
times\cite{tob92,Pacheco95,suisso02}. Considering the contact 
interaction  between the quarks, the 
baryon-Qqq vertex is described by two terms written 
as $v_{\alpha }(\vec{q}_{\perp },y)$ in the baryon rest frame,
 with $\alpha =\ q$ or $Q$, where $y=q^{+}/M_{B}$ is the Bjorken momentum
fraction and $M_B$ is the baryon mass. The separable structure
of the contact interaction implies that the vertex function depends 
only on the kinematical variables of the spectator quark in the
process of the interaction of the other two. As we are averaging over spin,
only two independent vertex functions are necessary to describe
the baryon wave function, i.e., $v_{q}$ and $v_{Q}$. 

%%%%%%%%%%%%%%%%%%%%%%%%%%%%%% FIG. 1 %%%%%%%%%%%%%%%%%%%%%%%%%%%%%

%% \begin{figure}[thbp]
%% \setlength{\epsfxsize}{0.8\hsize}
%% 
%% \special{psfile=fig1.eps   voffset=250  hoffset=15 
%% angle=-180 vscale=55   hscale=55}
%% \centerline{\epsfbox{fig1.eps}}
%\caption[dummy0]{
%Diagrammatic representation of Eq.\ref{Mass3}. The black 
%bubble represents the two-quark scattering amplitude.}
%% \label{fig1}
%% \end{figure}
%% {\small FIG. 1. Diagrammatic representation of Eq.\ref{Mass3}. The black 
%% bubble represents the two-quark scattering amplitude.} 
%%%%%%%%%%%%%%%%%%%%%%%%%%%%%%%%%%%%%%%%%%%%%%%%%%%%%%%%%%%%%%%%%%%

 The coupled Faddeev-Bethe-Salpeter equations for a heavy-light-light 
three quark system ($Qqq$) in the light-front are derived in the ladder 
approximation\cite{suisso02}, and are a generalization of the 
Weinberg equation\cite{Weinberg66} to three particle systems. Their
diagrammatical representation are given in figures 1 and 2. The first 
light-front equation in which the quark $Q$ is the spectator while 
the pair $qq$ interacts is represented in figure 1, from where one reads: 
\begin{eqnarray}
&&v_{Q}(\vec{q}_{\perp },y) =-\frac{2i}{\left( 2\pi \right) ^{3}}\tau
_{qq}\left( M_{qq}^{2}\right) \int\limits_{0}^{1-y}\frac{dx}{x\left(
1-x-y\right) }\int d^{2}k_{\perp }  \nonumber \\
&&\times \frac{\theta \left( x-\frac{m_{q}^{2}}{M_{B}^{2}}\right) \theta
\left( k_{\perp }^{\max }(m_{q})-k_{\perp }\right) 
v_{q}(\vec{k}_{\perp },x)}{M_{B}^{2}-\frac{%
q_{\perp }^{2}+m_{Q}^{2}}{y}-\frac{k_{\perp }^{2}+m_{q}^{2}}{x}-\frac{\left(
P_{B}-q-k\right) _{\perp }^{2}+m_{q}^{2}}{1-x-y}}.
\label{Mass3}
\end{eqnarray}
The second light-front equation, which is represented diagrammatically 
in figure 2, the quark pair $Qq$ interacts while the quark $q$ is the 
spectator. The second equation is given by:
\begin{eqnarray}
&&v_{q}(\vec{q}_{\perp },y) =-\frac{i}{\left( 2\pi \right) ^{3}}\tau
_{Qq}\left( M_{Qq}^{2}\right) \int\limits_{0}^{1-y}\frac{dx}{x\left(
1-x-y\right) }\int d^{2}k_{\perp }  \nonumber \\
&&\times \left[ \frac{\theta \left( x-\frac{m_{q}^{2}}{M_{B}^{2}}\right)
\theta \left( k_{\perp }^{\max }(m_{q})-k_{\perp }\right) 
v_{q}(\vec{k}_{\perp},x)}{M_{B}^{2}-\frac{%
q_{\perp }^{2}+m_{q}^{2}}{y}-\frac{k_{\perp }^{2}+m_{q}^{2}}{x}-\frac{\left(
P_{B}-q-k\right) _{\perp }^{2}+m_{Q}^{2}}{1-x-y}}\right.  \nonumber \\
&&\left. +\frac{\theta \left( x-\frac{m_{Q}^{2}}{M_{B}^{2}}\right) \theta
\left( k_{\perp }^{\max }(m_{Q})-k_{\perp }\right) 
v_{Q}(\vec{k}_{\perp },x)}{M_{B}^{2}-\frac{%
q_{\perp }^{2}+m_{q}^{2}}{y}-\frac{k_{\perp }^{2}+m_{Q}^{2}}{x}-\frac{\left(
P_{B}-q-k\right) _{\perp }^{2}+m_{q}^{2}}{1-x-y}}%
\right] \ .  \label{Mass32}
\end{eqnarray}
The maximum value for $k_{\perp }$ is
chosen to keep the mass squared of the $qq$ or $Qq$ subsystem real, i.e., $%
M_{qq}^{2}\geq 0$ and $M_{Qq}^{2}\geq 0$, respectively. These constraints in
the spectator quark phase-space come through the theta functions in the
integrations of Eqs. (\ref{Mass3}) and (\ref{Mass32}). For $M_{Qq}^{2}\geq 0$
one has $k_{\perp }<k_{\perp }^{\max }(m_{q})=\sqrt{%
(1-x)(M_{B}^{2}x-m_{q}^{2})}$, and $x\geq (m_{q}/M_{B})^{2}$. For $%
M_{qq}^{2}\geq 0$ one has $k_{\perp }<k_{\perp }^{\max }(m_{Q})=\sqrt{%
(1-x)(M_{B}^{2}x-m_{Q}^{2})}$, and $x\geq (m_{Q}/M_{B})^{2}$. For equal
particles, Eq.(\ref{Mass3}), reduces to the one derived in Ref.\cite{tob92} .
 The baryon four-momentum is given by $P_{B}$, the light and heavy quark
masses are $m_{q}$ and $m_{Q}$, respectively. The masses of the virtual
two-quark subsystems are $M_{qq}^{2}=(P_{B}-q)^{2}$ and $%
M_{Qq}^{2}=(P_{B}-q)^{2}$ due to the conservation of the total four-momentum.

%%%%%%%%%%%%%%%%%%%%%%%%%%%%%% FIG. 2 %%%%%%%%%%%%%%%%%%%%%%%%%%%%%
%% \begin{figure}[thbp]
%% \setlength{\epsfxsize}{1.\hsize}
%% \centerline{\epsfbox{fig2.eps}}
%\caption[dummy0]{
%Diagrammatic representation of Eq.\ref{Mass32}. The black bubble represents
%the two-quark scattering amplitude.}
%% \label{fig2}
%% \end{figure} 
%%%%%%%%%%%%%%%%%%%%%%%%%%%%%%%%%%%%%%%%%%%%%%%%%%%%%%%%%%%%%%%%%%%
%% {\small FIG. 2. Diagrammatic representation of Eq.\ref{Mass32}. 
%% The black bubble represents the two-quark scattering amplitude}
%%%%%%%%%%%%%%%%%%%%%%%%%%%%%%%%%%%%%%%%%%%%%%%%%%%%%%%%%%%%%%%%%%%%%%%%%%%

The two-quark scattering amplitudes $\tau _{qq}\left( M_{qq}^{2}\right) $ and 
$\tau _{Qq}\left( M_{Qq}^{2}\right) $ are the solutions of the
Bethe-Salpeter equations in the ladder approximation for a 
contact interaction between the quarks \cite{tob92,Pacheco2001}. In this 
approximation the scattering amplitude, which is the infinite sum of the 
powers of the product of the ``bubble''-diagram with the bare 
interaction strength, is given by the geometrical series: 
\begin{equation}
\tau _{\alpha q}\left( M_{\alpha q}^{2}\right) =\frac{1}{i\lambda ^{-1}-%
{\cal B}_{\alpha q}\left( M_{\alpha q}^{2}\right) }\ ,  \label{Mass8}
\end{equation}
where $\alpha =\ q$ or $Q$ and $\lambda $ is the bare interaction strength, and
\begin{eqnarray}
{\cal B}_{\alpha q}\left( M^2_{\alpha q}\right) &=&\int \frac{d^4k}{\left( 2\pi 
\right)^4}\frac{i}{(k^{2}-m_{q}^{2}+i\varepsilon) }
\nonumber \\
&\times &\frac{i}{(( P_{\alpha q}-k)
^{2}-m_{\alpha}^{2}+i \varepsilon)}  \ ,  
\label{Mass4}
\end{eqnarray} 
in which $P_{\alpha q}$ is the total four-momentum of the quark pair and 
$P_{\alpha q}^2=M_{\alpha q}^2$. 

The four-dimensional integration of the function 
${\cal B}_{\alpha q}\left(M_{\alpha q}^{2}\right) $ is 
the ``bubble''-diagram is performed in light-front variables. 
First, the virtual propagation of the
intermediate quarks is projected at equal light-front times \cite
{tob92,sales00}, by analytical integration over $k^{-}$ in the momentum
loop. Then, using the frame in which $\vec P_{\alpha q \perp}$ is zero
 and introducing the invariant quantity 
$x=\frac{k^{+}}{P_{\alpha q}^{+}}$, one obtains: 
\begin{eqnarray}
{\cal B}_{\alpha q}\left( M_{\alpha q}^{2}\right) &=& \frac{i}{2\left( 2\pi
\right) ^{3}}\int \frac{dxd^{2}k_{\perp }}{x\left( 1-x\right) }
\nonumber \\ &\times &
\frac{\theta(1-x)\theta (x)}{M_{\alpha q}^{2}
-\frac{k_{\perp }^{2}+\left( m_{\alpha
}^{2}-m_{q}^{2}\right) x+m_{q}^{2}}{x\left( 1-x\right) }}.  \label{Mass7}
\end{eqnarray}

We suppose that the light-quark pair system has a bound state, which allows 
to define the two-quark scattering amplitude. This physical condition
has been used in Ref. \cite{Pacheco95}. Therefore, 
the bound state pole of the light-quark scattering amplitude, $\tau
_{qq}(M_{qq}^{2})$ is found when $M_{qq}$ is equal to the mass of the bound $%
qq$ pair, $M_{d}$ which demands that $i\lambda ^{-1}={\cal B}_{qq}\left(
M_{d}^{2}\right)$. This is suficient to render finite the scattering
amplitudes $\tau _{qq}$ and $\tau _{Qq}$. Using that the bare strength
of the effective contact interaction between the constituent quarks $q$ and $%
Q$ does not depend on flavor, the final equation for the two-quark 
scattering amplitude is: 
\begin{equation}
\tau _{\alpha q}\left( M_{\alpha q}^{2}\right) =\frac{1}{{\cal B}_{qq}\left(
M_{d}^{2}\right) -{\cal B}_{\alpha q}\left( M_{\alpha q}^{2}\right) }\ .
\label{Mass11}
\end{equation}
The log-type divergence of $\tau _{\alpha q}$ is removed by the subtraction
in Eq.(\ref{Mass11}). 

\section{Results}

The solution of the coupled integral equations (\ref{Mass3}) 
and (\ref{Mass32}) for a relativistic system of three constituent 
quarks with a pairwise zero range interaction, gives the baryon mass
as a function of the quark mass $m_Q$. The physical inputs of the model are
the constituent quark mass $m_q$ and the diquark bound state mass. The
result for  the mass of the ground state baryon $(M_{B})$ alllows to calculate
the binding energy, defined by $B_{B}=2m_{q}+m_{Q}-M_{B}$. However,
the quarks are in fact confined in the hadron and to compare the model
with data, one has to define an experimental quantity which could be 
compared to the model binding energy. 

For that purpose, we use that the low-lying vector mesons are weakly bound
systems of constituent quarks while the pseudo-scalars are more strongly
bound within the same model\cite{tobpauli}. Therefore,
 we suppose that the masses of the constituent quarks
can be derived directly from the vector meson masses as:
\begin{eqnarray}
&&m_{u} =\frac{1}{2}M_{\rho }=0.384\ \text{GeV}\nonumber \\ 
&&m_{s}=M_{K^{\ast }}-\frac{1}{2}M_{\rho}=0.508\ \text{GeV} \nonumber  \\
&&m_{c}=M_{D^{\ast }}-\frac{1}{2}M_{\rho }=1.623\ \text{GeV}  \nonumber \\
&& m_{b}=M_{B^{\ast }}-\frac{1}{2}M_{\rho }=4.941\ \text{GeV},  
\label{mconst} 
\end{eqnarray}
where the masses of the mesons are taken from Ref.\cite{pdg}.

Using the constituent quark masses from Eq. (\ref{mconst}) and
the experimental values of the baryon masses \cite{pdg}, we can
attribute a binding energy to the low-lying spin 1/2 baryons, as given below: 
\begin{eqnarray}
&&B_{p}^{exp}=\frac{3}{2}M_{\rho }-M_{p}=0.214\ \text{GeV}
\nonumber \\
&&B_{\Lambda ^{0}}^{exp}=M_{K^{\ast}}+\frac{1}{2}M_{\rho }-M_{\Lambda ^{0}}
= 0.161\  \text{GeV}
\nonumber \\
&&B_{\Lambda _{c}^{+}}^{exp}=M_{D^{\ast }}+\frac{1}{2}M_{\rho }
-M_{\Lambda_{c}^{+}}=0.106 \ \text{GeV} \nonumber \\
&&B_{\Lambda _{b}^{0}}^{exp}=M_{B^{\ast }}+\frac{1}{2}M_{\rho
}-M_{\Lambda _{b}^{0}}=0.085 \ \text{GeV}\ .  
\label{bind}
\end{eqnarray}

%%%%%%%%%%%%%%%%%%%%%%%%%%%%%% FIG. 3 %%%%%%%%%%%%%%%%%%%%%%%%%%%%%
%% \begin{figure}[thbp]
%% \setlength{\epsfxsize}{0.8\hsize}
%% \centerline{\epsfbox{fig3.eps}}
%\caption[dummy0]{
%Low-lying hadron binding energy $(B)$ of the low-lying hadrons 
%as a function of the corresponding ground state mass. Experimental data
%for pseudoscalar mesons from Table I\cite{pdg} (full squares). 
%Experimental data for the low-lying spin 1/2 baryons comes from
%Table II\cite{pdg} (empty circles). The results of the light-front
%model from the solution of Eqs.\ref{Mass3} and \ref{Mass32} are 
%shown by solid line.}
%% \label{fig3}
%% \end{figure}
%% {\small FIG. 3. Low-lying hadron binding energy 
%% $(B)$ of the low-lying hadrons 
%% as a function of the corresponding ground state mass. Experimental data
%% for pseudoscalar mesons from Table I\cite{pdg} (full squares). 
%% Experimental data for the low-lying spin 1/2 baryons comes from
%% Table II\cite{pdg} (empty circles). The results of the light-front
%% model from the solution of Eqs.\ref{Mass3} and \ref{Mass32} are 
%% shown by solid line.} 
%%%%%%%%%%%%%%%%%%%%%%%%%%%%%%%%%%%%%%%%%%%%%%%%%%%%%%%%%%%%%%%%%%%

In figure 3, we plot the binding energies of the low-lying pseudo-scalar
mesons, defined as $B_{M}=M_{v}-M_{ps}$, where $M_v$ and $M_{ps}$ are
the masses of the vector and pseudoscalar low-lying mesons respectively,
 against the mass of the corresponding
pseudo-scalar meson. Also in the figure is shown 
the values of the binding energies of the spin
1/2 baryons ($N$, $\Lambda ^{0}$, $\Lambda _{c}^{+}$ and $\Lambda _{b}^{0}$)
from Eq. (\ref{bind}) as a function of the corresponding baryon
mass. The systematic behaviour of the defined binding energy
for the hadron is qualitative the same independent of the quark content.

The results from the numerical solution of  the coupled equations 
(\ref{Mass3}) and (\ref{Mass32}) are obtained for a fixed $m_{u}=0.386$ GeV 
(we kept the value found in Ref.\cite{Pacheco95}) which together with 
the nucleon mass of $0.938$ GeV implies in $M_{d}=0.695$ GeV. For the
given value of the diquark mass $M_d$ and  different values of $m_{Q}$, we
obtain  the binding
energy for the spin 1/2 baryons $\Lambda ^{0}$, $\Lambda _{c}^{+}$ and $%
\Lambda _{b}^{0}$ as a function of $m_{Q}$ ($Q=s,c,b$). 
For the baryon
mass above $2.3$ GeV,the bound $Qqq$ 
system of the light-front model goes to the diquark threshold.
This gives the saturation value of $0.077GeV$ seen in this figure.
The theoretical results are in
excelent agreement with the baryon data, consequently the dynamical
assumption of the flavor independence of the effective interaction is indeed
reasonable. In figure 4, we shown the baryon binding energy as a function of $%
m_{Q}$, again we observe the agreement between the model and  
the attributed experimental values for the binding energies and constituent
quark masses.

%%%%%%%%%%%%%%%%%%%%%%%%%%%%%% FIG. 4 %%%%%%%%%%%%%%%%%%%%%%%%%%%%%%%%%%%%%% 
%% \begin{figure}[thbp]
%% \setlength{\epsfxsize}{0.8\hsize}
%% \centerline{\epsfbox{fig4.eps}}
%\caption[dummy0]{
%Low-lying baryon binding energy $(B)$ 
%as a function of the constituent quark mass $(M_Q)$. 
%Experimental data for the low-lying spin 1/2 baryons comes from 
%Table II\cite{pdg} (empty circles) and the constituent quark masses are given
%in Table I. The results of the light-front
%model from the solution of Eqs.\ref{Mass3} and \ref{Mass32} are 
%shown by solid line.}
%% \label{fig4}
%% \end{figure}  
%% {\small FIG. 4. Low-lying baryon binding energy $(B)$ 
%% as a function of the constituent quark mass $(M_Q)$. 
%% Experimental data for the low-lying spin 1/2 baryons comes from 
%% Table II\cite{pdg} (empty circles) and the constituent quark 
%% masses are given
%% in Table I. The results of the light-front
%% model from the solution of Eqs.\ref{Mass3} and \ref{Mass32} are 
%% shown by solid line.}
%%%%%%%%%%%%%%%%%%%%%%%%%%%%%%%%%%%%%%%%%%%%%%%%%%%%%%%%%%%%%%%%%%%%%%%%% 

\section{Conclusions}

The binding energy of the constituent quarks forming the the
low-lying spin 1/2 baryonic states of the nucleon, $\Lambda ^{0}$, $\Lambda
_{c}^{+}$ and $\Lambda _{b}^{0}$, obtained from the experimental values of
baryon masses and constituent quark masses derived from 
the low-lying vector mesons masses was studied within a light-front
model. The effective interaction between the constituent quarks 
was chosen of a contact form and spin was averaged out. The motivation of
chosing this particular interaction was two-fold: it was necessary to
bring to the effective QCD-inspired model of the light-mesons the pion
mass scale, and to give the observed splitting between the pion and 
rho meson spectrum\cite{tobpauli}, on one side, and on the other 
side, it was succesfull in describing the proton mass and radius 
simultaneously\cite{Pacheco95}. In the present study the
contact interaction was used because it allows to introduce in the model
the minimal number of physical scales necessary to 
describe the low-lying spin 1/2 baryons. Therefore, 
the relativistic three-quark model of the baryon defined on the 
light-front, with a falvor independent interaction, has as inputs  
the constituent quark masses and the diquark mass in the light 
sector, which defines the strength of the contact interaction. 
This model allowed a surprising reproduction of the trend and 
magnitude of the binding energies as a function of the distint quark mass.
As the present light-front model is still very schematic, we believe that 
our conclusion of the flavor dependence of baryonic 
masses may still hold in a more realistic model, and support the extension
of the QCD-inspired model applied previously only to mesons\cite{pauli,tobpauli}
also to baryons.

Acknowlegments: EFS and JPBCM thank the Brazilian funding agencies FAPESP
(Funda\c{c}\~{a}o de Amparo a Pesquisa do Estado de S\~{a}o Paulo) and TF
thanks FAPESP and CNPq (Conselho Nacional de Pesquisa e Desenvolvimento of
Brazil).

%% \newpage 

%% .

%% FIGURES 

%%%%%%%%%%%%%%%%%%%%%%%%%%%%%% FIG. 1 %%%%%%%%%%%%%%%%%%%%%%%%%%%%%

\begin{figure}[h]
\vspace{10.0cm}
\includegraphics{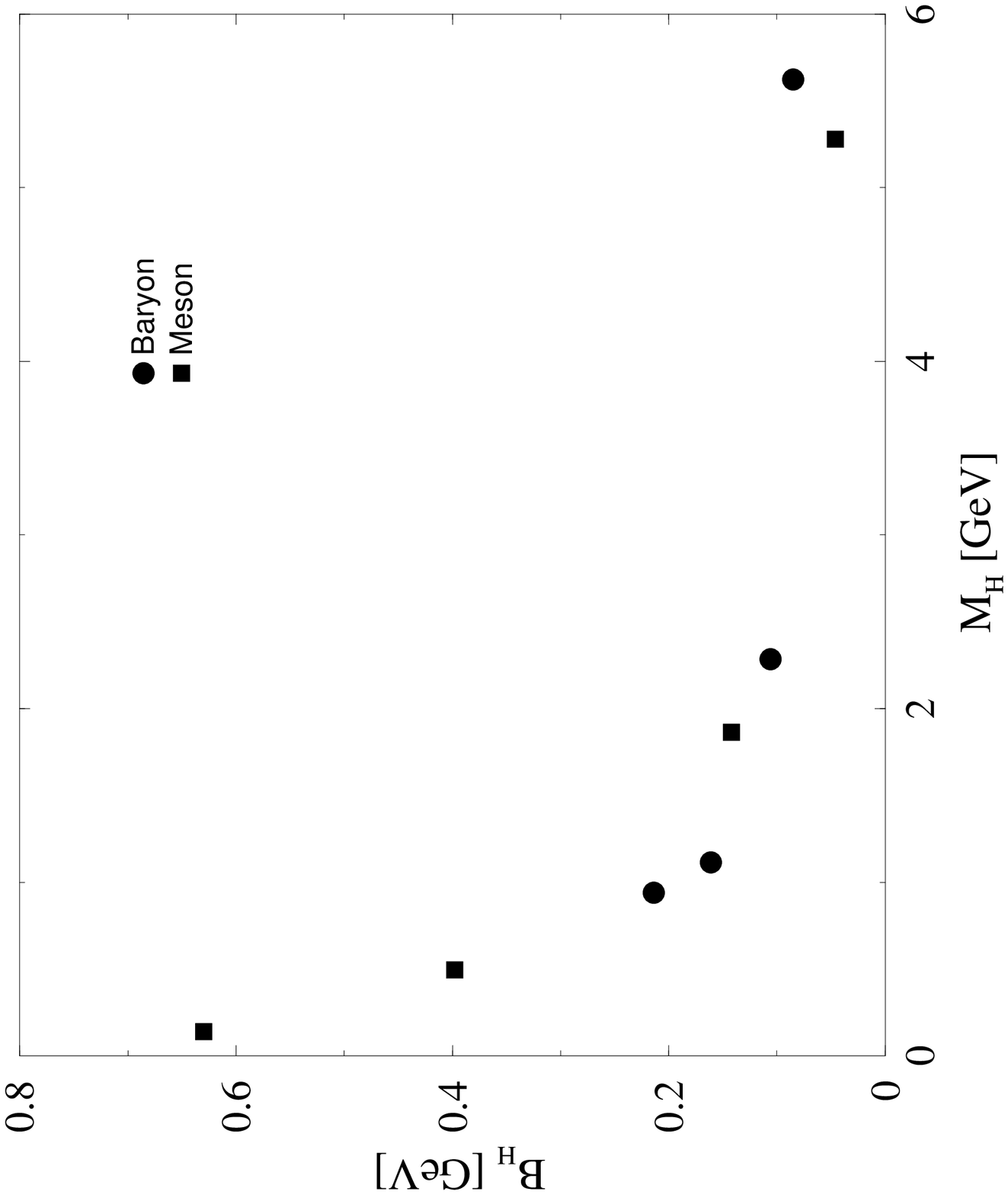}
\vspace{3.0cm}
\caption{Diagrammatic representation of Eq.\ref{Mass3}. The black 
bubble represents the two-quark scattering amplitude.} 
\label{vnl1s0}
\vspace{1.2cm}
\end{figure}

%%%%%%%%%%%%%%%%%%%%%%%%%%%%%% FIG. 2 %%%%%%%%%%%%%%%%%%%%%%%%%%%%%
\begin{figure}[thbp]
\setlength{\epsfxsize}{1.\hsize}
\centerline{\epsfbox{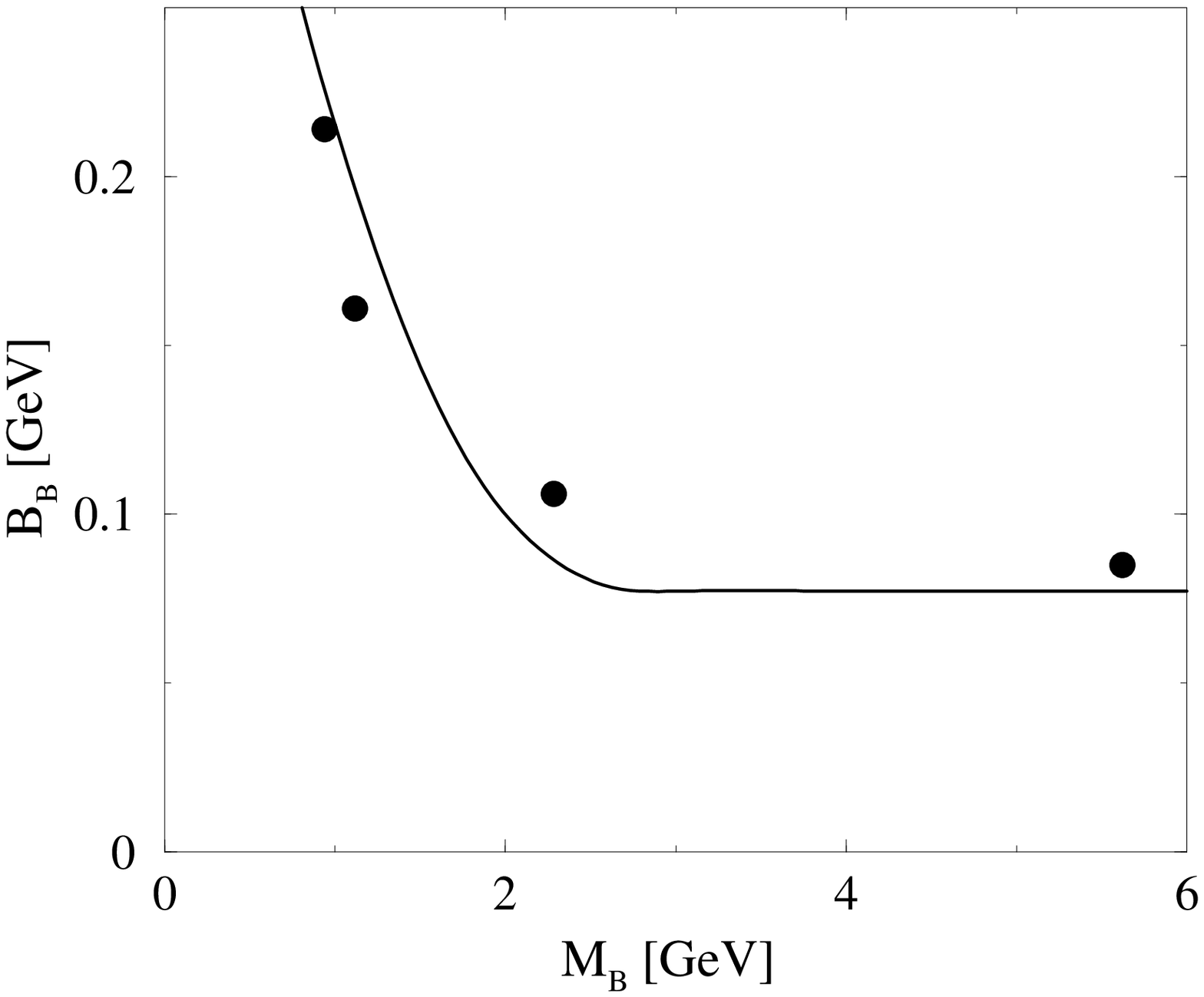}}
%\caption[dummy0]{
%Diagrammatic representation of Eq.\ref{Mass32}. The black bubble represents
%the two-quark scattering amplitude.}
\label{fig2}
\end{figure} 
%%%%%%%%%%%%%%%%%%%%%%%%%%%%%%%%%%%%%%%%%%%%%%%%%%%%%%%%%%%%%%%%%%%
{\small FIG. 2. Diagrammatic representation of Eq.\ref{Mass32}. 
The black bubble represents the two-quark scattering amplitude}

%%% \newpage 

%%%%%%%%%%%%%%%%%%%%%%%%%%%%%% FIG. 3 %%%%%%%%%%%%%%%%%%%%%%%%%%%%%
\begin{figure}[thbp]
\setlength{\epsfxsize}{0.8\hsize}
\centerline{\epsfbox{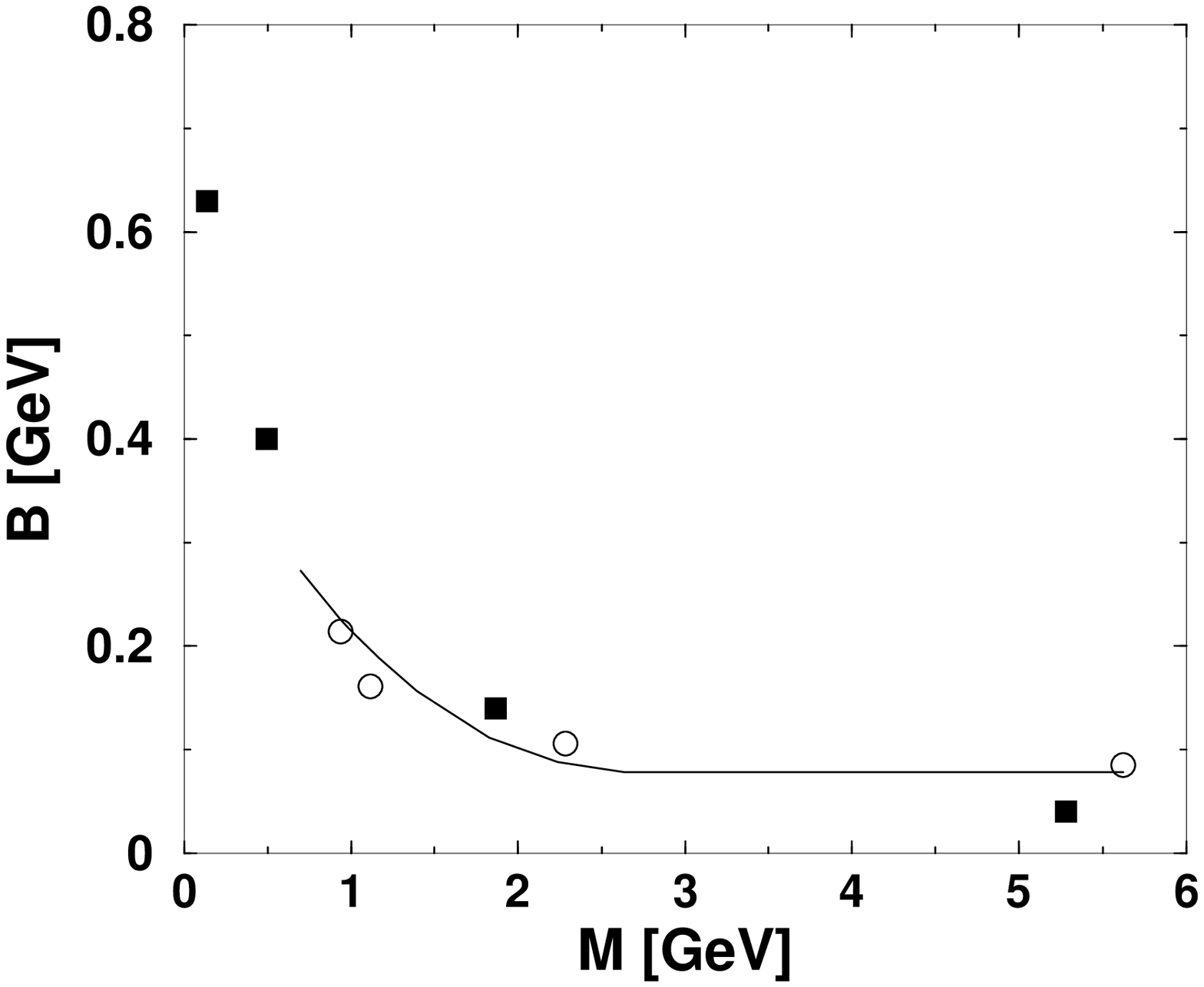}}
%\caption[dummy0]{
%Low-lying hadron binding energy $(B)$ of the low-lying hadrons 
%as a function of the corresponding ground state mass. Experimental data
%for pseudoscalar mesons from Table I\cite{pdg} (full squares). 
%Experimental data for the low-lying spin 1/2 baryons comes from
%Table II\cite{pdg} (empty circles). The results of the light-front
%model from the solution of Eqs.\ref{Mass3} and \ref{Mass32} are 
%shown by solid line.}
\label{fig3}
\end{figure}
{\small FIG. 3. Low-lying hadron binding energy $(B)$ of the low-lying hadrons 
as a function of the corresponding ground state mass. Experimental data
for pseudoscalar mesons from Table I\cite{pdg} (full squares). 
Experimental data for the low-lying spin 1/2 baryons comes from
Table II\cite{pdg} (empty circles). The results of the light-front
model from the solution of Eqs.\ref{Mass3} and \ref{Mass32} are 
shown by solid line.}

%%%%%%%%%%%%%%%%%%%%%%%%%%%%%% FIG. 4 %%%%%%%%%%%%%%%%%%%%%%%%%%%%%%%%%%%%%% 
\begin{figure}[thbp]
\setlength{\epsfxsize}{0.8\hsize}
\centerline{\epsfbox{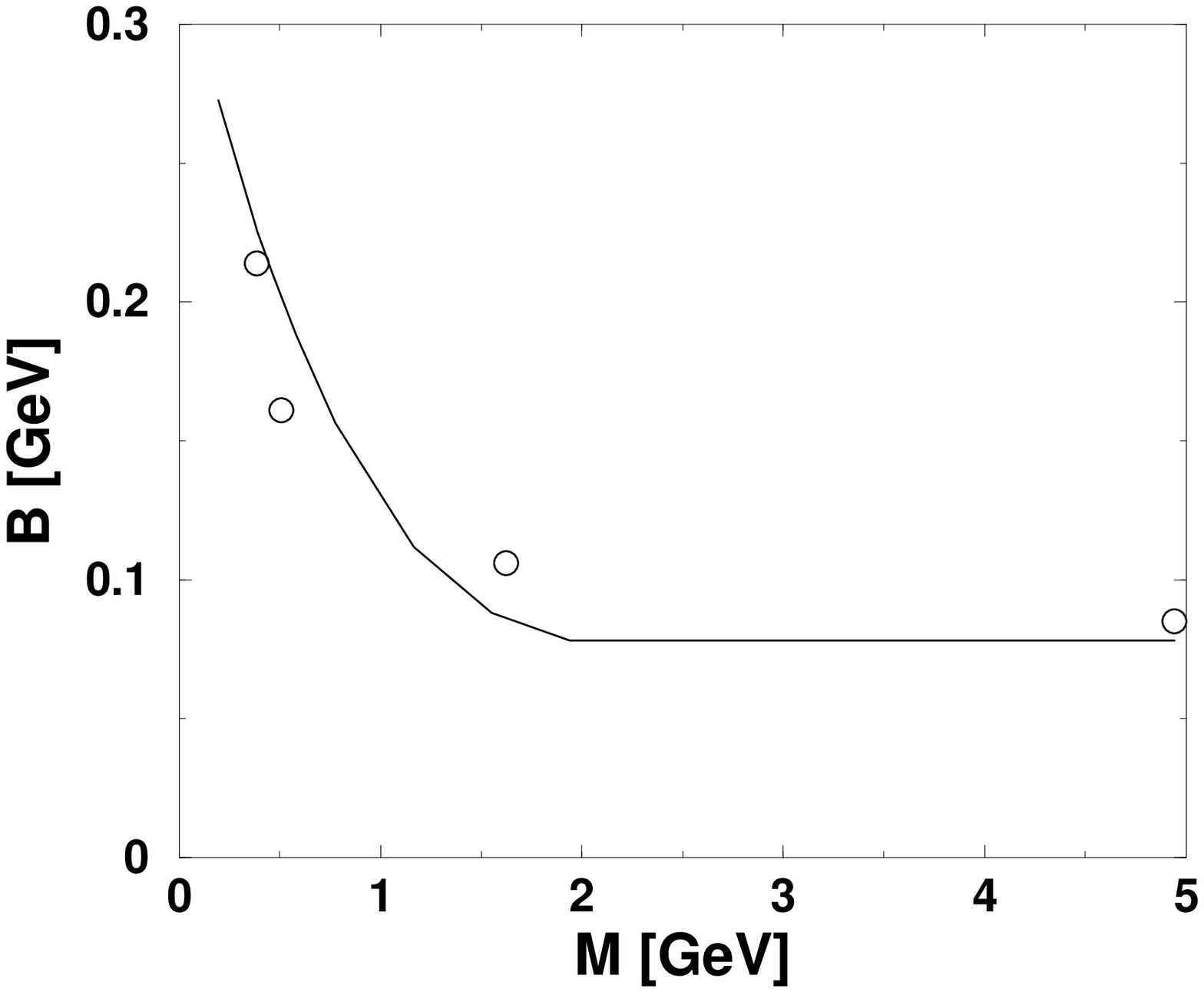}}
%\caption[dummy0]{
%Low-lying baryon binding energy $(B)$ 
%as a function of the constituent quark mass $(M_Q)$. 
%Experimental data for the low-lying spin 1/2 baryons comes from 
%Table II\cite{pdg} (empty circles) and the constituent quark masses are given
%in Table I. The results of the light-front
%model from the solution of Eqs.\ref{Mass3} and \ref{Mass32} are 
%shown by solid line.}
\label{fig4}
\end{figure} 
{\small FIG. 4. Low-lying baryon binding energy $(B)$ 
as a function of the constituent quark mass $(M_Q)$. 
Experimental data for the low-lying spin 1/2 baryons comes from 
Table II\cite{pdg} (empty circles) and the constituent quark masses are given
in Table I. The results of the light-front
model from the solution of Eqs.\ref{Mass3} and \ref{Mass32} are 
shown by solid line.}
%%%%%%%%%%%%%%%%%%%%%%%%%%%%%%%%%%%%%%%%%%%%%%%%%%%%%%%%%%%%%%%%%%%%%%%%% 

%% \end{multicols}

\end{document}